\documentclass[aps,prb,twocolumn,showpacs,amsmath,amssymb,superscriptaddress, citeautoscript]{revtex4-2}
\setcitestyle{super}
\usepackage[cmtip,all]{xy}
\usepackage{graphicx}
\usepackage{hyperref}
\usepackage{xcolor}
\usepackage{color}
\usepackage[all,cmtip]{xy}
\usepackage{multirow}
\usepackage{natbib}
\usepackage{amsmath,amsthm} 
\usepackage{amsfonts}

\makeatother

\begin{document}

\title{Spin Chern number in altermagnets}

\author{Rafael Gonz\'{a}lez-Hern\'{a}ndez}
\email{rhernandezj@uninorte.edu.co}
\affiliation{Departamento de F\'{i}sica y Geociencias, Universidad del Norte, Km. 5 V\'{i}a Antigua Puerto Colombia, Barranquilla 081007, Colombia}
\author{Higinio Serrano}
\email{hserrano@math.cinvestav.mx}
\affiliation{Centro de Investigaciones y Estudios Avanzados, Av. Instituto Politécnico Nacional 2508,
Col. San Pedro Zacatenco, Mexico D.F., CP 07360, Mexico}
\author{Bernardo Uribe}
\email{bjongbloed@uninorte.edu.co}
\affiliation{Departamento de Matem\'{a}ticas y Estad\'{i}stica, Universidad del Norte, Km. 5 V\'{i}a Antigua Puerto Colombia, Barranquilla 081007, Colombia}

\date{\today}

\begin{abstract}
	This work explores the topological properties of altermagnets, a novel class of collinear magnetic materials.
	We employ equivariant K-theory of magnetic groups and Hamiltonian models to formulate a robust $C^z_4 \mathbb{T}$ topological invariant to classify 2D and 3D altermagnetic systems. 
	Our findings demonstrate that the spin Chern number serves as a robust topological index, corresponding to the half-quantized Chern number of the divided Brillouin zone. 
	This indicator enables the prediction of a topologically protected 2D altermagnetic insulators and 3D Weyl altermagnetic semimetals, highlighting the relationship between altermagnetism and topological phases.
	Furthermore, our results provide a pathway to the exploration of topological applications in $d$-wave altermagnetic materials.
\end{abstract}
\maketitle

\section*{Introduction}

Altermagnetism \cite{Smejkal2020,Smejkal2022,Smejkal2022_2,smejkal_nat_review_22,Kusunose2019,Kunes2019} has emerged as a novel class of collinear magnetic phase distinguished by a unique breaking of time-reversal symmetry ($\mathbb{T}$) in reciprocal space, despite of having a compensated magnetic order in real space. 
Unlike conventional collinear antiferromagnets (AFMs), altermagnets exhibit a distinct electronic band structure due to their sublattices of opposite spin being connected not through simple translation or inversion, but through non-trivial rotations \cite{Turek2022,Bai2023,Smejkal2024,jiang2024}.
For example, the combination of four-fold rotation with time-reversal symmetry ($C^z_4 \mathbb{T}$) was identified in the metallic RuO$_2$, this one being  the first $d$-wave altermagnetic material discovered \cite{Smejkal2020,Smejkal2022chiral}.
In this system, anisotropic spin charge densities lead to considerable spin-splitting in the electronic bands, which is independent of relativistic spin-orbit coupling (SOC) interactions. 
This spin-splitting produce unconventional magnetic responses, such as the anomalous Hall \cite{Feng2022,Betancourt2023,intrinsic-ahe2024}, spin-filter \cite{Gonzalez2021,PhysRevX2022} and the magneto-optical Kerr effect \cite{Isaiah2024}, which make altermagnets an interesting topic for both theoretical ~\cite{PhysRevX.12.040002,mcclarty2023landau,fesb2-am,PhysRevB.105.064430,PhysRevB.102.144441,PhysRevB.108.184505,PhysRevLett.132.056701,Jaeschke2023,Zhu2023multipiezo}
and experimental 
~\cite{zhu_nature_24,krempasky_jungwirth_nature_24,fedchenko_sci_advances_24,Reichlova2024,lin2024observation} investigation in condensed matter physics.

On the other hand, the topological materials can be classified into categories such as topological insulators and topological semimetals \cite{hasan2010colloquium,qi2010topological,RevModPhys.88.035005}.
The latter can defined by the presence of symmetry protected Weyl points (WPs) at the Fermi level.
These WPs usually appear when spin degeneracy is lifted through mechanisms like strong SOC in inversion or time-reversal symmetry breaking systems \cite{Topological-semimetals,Phases-Weyl-semimetals}.
While traditional AFMs cannot support WPs due to their spin-degenerate bands, altermagnets are expected to exhibit WPs naturally due to their spin-split bands \cite{Smejkal2022,Smejkal2022_2}.
These unique properties have generated interest in the combination of altermagnetic and topological properties which are protected by symmetry operations. 

However, defining topological invariants in insulating or semimetallic altermagnets poses a challenge, particularly due to the absence of time-reversal symmetry, 
which renders the $\mathbb{Z}_2$ index undefined \cite{Fu-Kane,Fu-Kane-Mele}.
Additionally, the $C^z_4 \mathbb{T}$ symmetry causes the Chern number to vanish in the entire Brillouin zone (BZ), which also restricts its potential as a possible topological indicator \cite{TKNN-invariant,Berry-phase}.  
Recently, the $C^z_4 \mathbb{T}$ symmetry has been employed to divide the BZ into two regions, allowing for the calculation of a half-quantized Chern number with opposite signs \cite{C4T_2024}.

In this work, we have employed equivariant K-theory of magnetic groups to establish a robust $C^z_4 \mathbb{T}$  topological invariant for characterizing altermagnetic systems. 
Using K-theory and Hamiltonian models, we demonstrates that the spin Chern number (SCN) \cite{Prodan-SCN} can be a topological index for classifying 2D and 3D insulating and semimetallic altermagnets with $C^z_4 \mathbb{T}$ symmetry-protected states. 
We prove that the SCN corresponds to the half-quantized Chern number assigned to the divided BZ, and therefore the SCN becomes a clear indicator to identify topological altermagnetic phases.
Finally, the topologically protected 2D insulator and 3D Weyl semimetal altermagnets are predicted through spin topology characterization \cite{Gonzalez-Uribe,spin-resolved-3d,PhysRevResearch.5.033013}. 

\section{2D altermagnetic topological insulator}

\begin{figure*}
	\includegraphics[width=18cm]{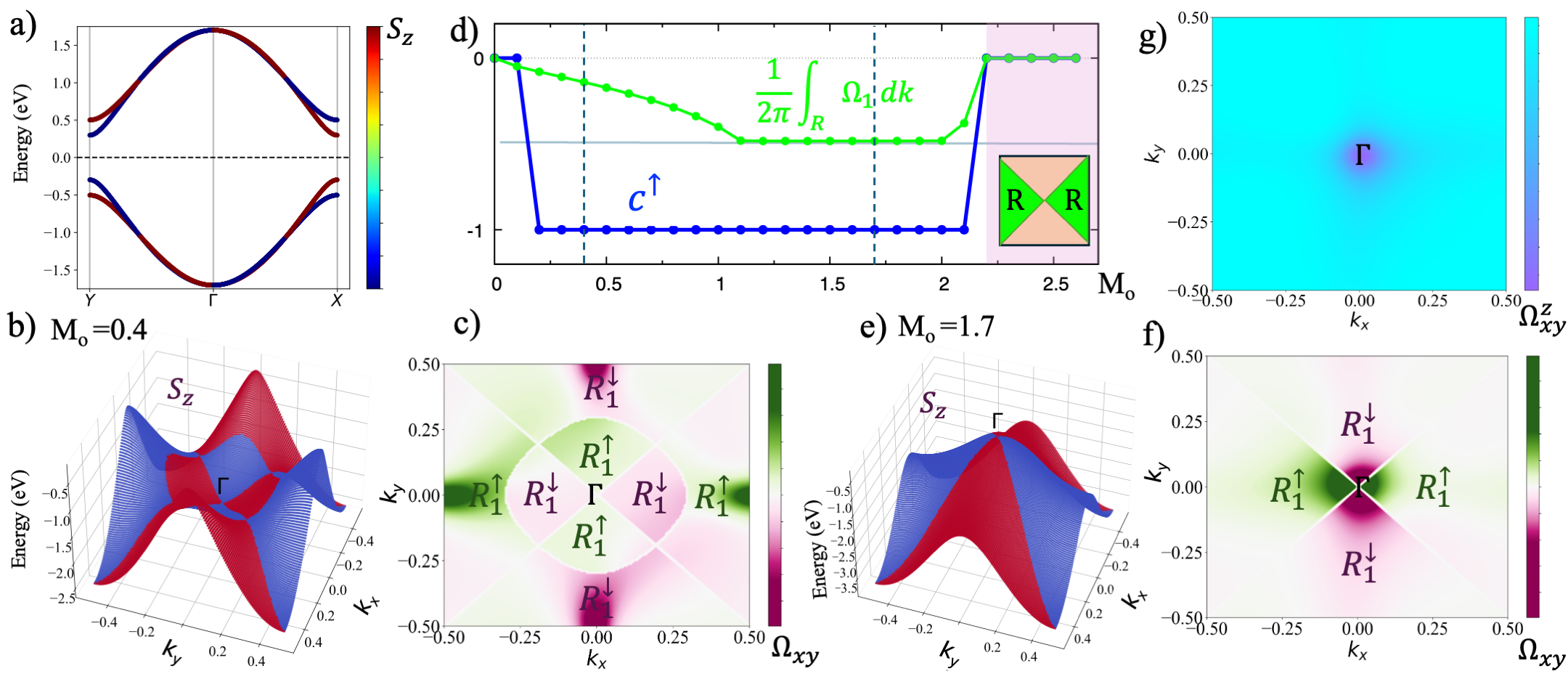}
	\caption{(a) Spin splitting of the electronic band structure of the two-dimensional (2D) altermagnet topological insulator (TI) derived from the Hamiltonian in Eqn. \eqref{Altermagnetic_Hamiltonian}, with $G$=1.05. (b) Spin projection at the top of the valence band and (c) Berry curvature for the 2D altermagnet TI with a mass term $M_0 = 0.4$. (d) Phase diagram of spin-up Chern number $c^\uparrow$ and the integral of Berry curvature $\Omega_1$ over half of the Brillouin zone $R$, as shown in the panel, as a function of the mass term $M_0$. Non-trivial and trivial phases are classified with the variation of the Chern numbers.  Note that the value of the Chern number depends on the radius of the nodal line for $M_0 < G^2$.  (e) Spin projection at the top of the valence band ($M_0 = 1.7$), illustrating the nodal lines at $k_x = \pm k_y$ where a change in spin value occurs. (f) Berry curvature and (g) Spin Berry curvature for the case of $M_0 = 1.7$. All the Hamiltonian parameters are in $eV$.  The integration regions $R_1^\uparrow$ and $R_1^\downarrow$, used in the calculations of $c_1^{\uparrow}$ and $c_1^{\uparrow}$ in Eqn. \ref{scn_up}, are shown in panels (c) and (f). The Fermi energy is set to zero.}
	\label{figure 2D}
\end{figure*}

We consider a four band model on a square lattice in which unit-cell contains the following orbitals: $d_{xy \uparrow}$, $\frac{1}{\sqrt{2}}(d_{xz \uparrow}-id_{yz \uparrow})$, $d_{xy \downarrow}$ and $\frac{1}{\sqrt{2}}(d_{xz \downarrow}-id_{yz \downarrow})$. 
In this model, a 2D altermagnetic Hamiltonian similar to the Bernevig-Hughes-Zhang (BHZ) \cite{BHZ} in momentum space, has been proposed by Ma and Jia\cite{Altermagnetic-C4T}, taking the following form:
\begin{align}
H = 
\left(\begin{matrix}
M_1 & A+iGB & 0 & 0 \\
A-iGB & -M_1 & 0 & 0 \\
0 & 0 & M_2 & B+iGA \\
0 & 0 & B-iGA & -M_2
\end{matrix} \right). \label{Altermagnetic_Hamiltonian}
\end{align}
with
\begin{align}
A =& A_0 \sin(k_x) \\
B =& B_0 \sin(k_y) \\
M_1 =& M_0 - (\cos(k_x) +G^2 \cos(k_y))\\
M_2 =& M_0 - (G^2\cos(k_x) + \cos(k_y)).
\end{align}

Here, $M_1$ and $M_2$ represent the mass terms for each spin channel, while $M_0$ denotes the on-site atomic potential which can vary with the value of the local magnetic moment.
The parameters $A_0$ and $B_0$ describe the hopping amplitudes between $d$ orbitals and we will set these to $1$ eV for convenience.
The constant $G$ introduces a symmetry breaking between the $x$ and $y$ directions within each spin block, making the Hamiltonian characteristic of a $d$-wave altermagnetic 2D system when G $\neq$ 1  \cite{Altermagnetic-C4T}.
We set the constant $G$ to be a number slightly shifted from $1$ (in $eV$):
\begin{align}
G\sim 1.
\end{align}
This Hamiltonian  preserves both the $C^z_4 \mathbb{T}$  and the inversion  $I$ symmetries with matrix operators:
\begin{align}
C_4^z\mathbb{T} =
\left(\begin{matrix}
0 & 0& i& 0 \\
0 & 0 & 0 & -i  \\
-1 & 0 & 0 & 0 \\
0 & -1 & 0 & 0
\end{matrix} \right) \mathbb{K}, \ \ 
I =\mathrm{diag}(1,-1,1,-1)
\end{align}
where we have  $C_4^z\mathbb{T} (k_x,k_y) = (k_y,-k_x)$.

The spin $S_z$ operator in this basis is given by the diagonal matrix
\begin{align}
S_z =  \mathrm{diag}(1,1,-1,-1)
\end{align}
and it commutes with the Hamiltonian.

The energies of the valence bands are 
\begin{align}
\min(\lambda_1,\lambda_2) \ \ \mbox{and} \ \ \max(\lambda_1,\lambda_2)
\end{align}
where 
\begin{align}
\lambda_1 = - \sqrt{A^2 + G^2B^2 + M_1^2} \\
\lambda_2 = - \sqrt{G^2 A^2 + B^2 + M_2^2}.
\end{align}
The Hamiltonian is energy gapped unless $A_0=0=B_0$ and either $M_1=0$ or $M_2=0$. The values for $M_0$ that
close the energy gap are given by $M_0= \pm 1 \pm G$. Denote the eigenstates of the Hamiltonian by $\psi_0$ and $\psi_1$.

Since the spin ($S_z$) is a good quantum number, the valence bands split into spin-up and spin-down components, exhibiting opposite spin splittings for the $x$  and $y$ directions in the first Brillouin zone (BZ), as it is shown in the Fig \ref{figure 2D}a.  

This spin splitting is given by the following eigenvectors of the Hamiltonian:

\begin{align}
\nu^\uparrow &= \langle M_1+\lambda_1,A-iGB,0,0 \rangle   \label{spin up in H}   \\
\nu^\downarrow &= \langle0,0, M_2+\lambda_2,B-iGA\rangle.  \label{spin down in H}
\end{align}

Using these eigenvectors, we have calculated the Berry curvature for the spin-up states and found that the Chern number of the spin-up band, which is the opposite of the spin-down band. 
The Chern numbers are as follows:

\begin{align} 
c_1(\nu^\uparrow) = - c_1(\nu^\downarrow) = 
\left\{ \begin{matrix}
1 & \mbox{for} & G-1<M_0<1+G\\
-1 & \mbox{for} & -1-G < M_0 <1-G \\
0 & \mbox{for} & 1+G<|M_0|.
\end{matrix} \right.
\end{align}

Due to the spin character changing inside each of the valence bands, as shown in Fig. \ref{figure 2D}e, this spin texture induce nodal lines in the altermagnetic systems, as predicted by several works \cite{nodalines-am,PhysRevLett.133.146602,topologicalresponsesgappedweyl,mirrorchernbandsweyl}. 
The valence bands intersect along the nodal lines at $k_x=\pm k_y$ in reciprocal space. When no additional intersections occur, the upper and lower valence bands split into two connected regions, as depicted in Fig. \ref{figure 2D}e. 
This is the case when $|M_0|>G^2$.

For the case of $|M_0 |< G^2$ the two valence bands further intersect on a
nodal line around the $\Gamma$ point as depicted in Figs. \ref{figure  2D} b) \& c), crossing the $k_x$ axis on the points with coordinates

\begin{align}
\left(\pm\cos^{-1}\left(1-\tfrac{2M_0}{G^2}\right),0\right).
\end{align}

On the other hand, the $C^z_4 \mathbb{T}$ symmetry forces the Chern number of the Bloch bundle of valence bands to be zero. 
This arises because the Berry curvature alternates between equivalent positive and negative values in the BZ, as shown in Figs. \ref{figure 2D}c and \ref{figure 2D}f, making the total integral of the Berry curvature zero. As a result, the system does not exhibit an anomalous Hall response.

However, the spin structure of the upper valence band allows us to compute the integral of the Berry curvature tensor over the region where the band is exclusively spin-up.
We define two distinct regions in the BZ, which are defined by the $C^z_4 \mathbb{T}$ symmetry:

\begin{align}
R^{\uparrow, \downarrow}_j = \{ {\bf{k}} \colon \langle \psi_j({\bf{k}})|S_z  \psi_j({\bf{k}}) \rangle = \pm 1 \},
\end{align}
we note that $R_0^\downarrow = R_1^\uparrow$,  $  R_0^\uparrow=R_1^\downarrow$ and $C_4R_1^\uparrow =R_1^\downarrow$.

Each region is half of the BZ and they can be seen in Figs. \ref{figure  2D} c) and f) for different values of $M_0$ (0.4 and 1.7 $eV$). 

If $\Omega_j$ denotes the Berry curvature of the $j$-th valence band $\psi_j$, we have that 
\begin{align}
 \tfrac{1}{2}c_1^{\uparrow} =  \tfrac{1}{2\pi}\int_{R_1^\uparrow} \Omega_1 d {\bf{k}} =-
\tfrac{1}{2\pi}\int_{R_1^\downarrow} \Omega_1 d {\bf{k}}.
\label{scn_up}
\end{align}

Therefore the integral of the Berry curvature for each spin region also permits to determine the spin Chern  number.
To calculate the Chern and spin Chern numbers, we have constructed the Hamiltonian model using the \textsl{pythtb} code \cite{pythtb}.

In the case that the valence bands only intersect on
the lines $k_x=\pm k_y$, in other words when $|M_0| > G^2$, the regions defined above are squares. 
These squares are depicted in Fig. \ref{figure  2D} f).

Integrating the Berry curvature of the upper valence band on this region and varying $M_0$  we obtain the phase diagram of the 
Fig. \ref{figure  2D} d). 
Here can be noticed that for $0<M_0<G^2$ there is a nodal line in the valence bands around the $\Gamma$ point,
and therefore the integral of the Berry curvature on the squares of Fig. \ref{figure  2D} f) is not $\tfrac{1}{2}$. 
The value of this integral starts at $0$, for $M_0=0$ and increases to $\tfrac{1}{2}$ when $M_0=G^2$. 

For values of $M_0$ bigger than $G^2$, the valence bands separate,  except on the lines $k_x=\pm k_y$, and the value of the integral of the Berry curvature is $\tfrac{1}{2}$.  
This result is consistent with the half-quantized Berry curvature predicted for two-dimensional  $C_4^z\mathbb{T}$-symmetric topological semimetals \cite{C4T_2024}.

On the interval $G^2 < M_0 < 1+G$ the integral of the Berry curvature possess the same information as the value of the 
spin-up Chern number, which we have calculated separately.
However, the spin-up Chern number in the phase diagram shown in Fig. 1d) indicates a topological transition from $c_1(\nu^\uparrow) = -1$ to $0$ at $M_0 \sim 0$ and $M_0 = 1 + G^2$. 
At these values of $M_0$, the system becomes gapless, resulting in the emergence of edge states when two distinct topological phases are in contact.
 
The topological character of this 2D altermagnet can be distinguished by measuring the quantized spin Hall response, which is determined by the spin Chern number. 
As we have demonstrated, the spin Chern number can be a robust topological invariant for characterizing 2D altermagnetic phases with $C_4^z\mathbb{T}$ symmetry, which allows a clean separation of the spin channels for each valence bands. 
This invariant can be used to predict the topological phases of the $d$-wave 2D altermagnetic materials, such as FeSe\cite{mazin-fese}, as well as those recently proposed by Bai \textit{et al.} \cite{Smejkal2024}

 \section{3D Weyl semimetal altermagnet}

\begin{figure*}
	\includegraphics[width=18cm]{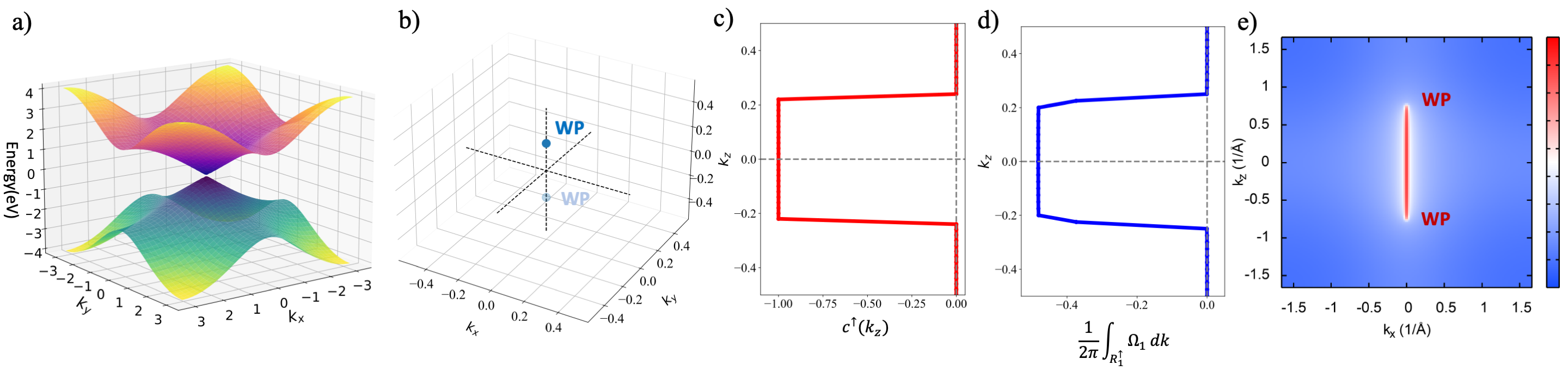}
	\caption{(a) 2D Energy band structure of the three-dimensional (3D) Weyl semimetal altermagnet, based on the Hamiltonian in Eqn. \eqref{Altermagnetic_Hamiltonian_3D}, with $G$=1.05 and $M_0$= 2.2, for the $k_z$=$0.22$ plane where the Weyl points (WPs) are located.  The negligible SOC  parameters in the Hamiltonian \eqref{Altermagnetic_Hamiltonian_3D} are set $C_0$=$0$, $D_0$=$0$, (b) Locations of the WPs in reciprocal space, which are found along the $k_z$ axis. (c) Spin-up Chern number $c^\uparrow$ and the integral of Berry curvature $\Omega_1$ for the region $R_1^\uparrow$ as a function of the $k_z$ plane within the 3D altermagnetic phase. It is noteworthy that the transition of the Chern numbers occurs when the $k_z$ planes cross in the WPs, which provide topological protection. (d) Electronic surface states for the $k_x$-$k_z$ plane of the altermagnetic 3D model, with the Fermi arc showing the connection between opposite WP chiralities.}
	\label{figure2}
\end{figure*}

In order to upgrade the 2D Hamiltonian of Eqn. \eqref{Altermagnetic_Hamiltonian} to a 3D one, 
we include electronic interactions along the $z$-axis into the mass term
\begin{align}
M_1 =& M_0 - (\cos(k_x) +G^2 \cos(k_y) + \cos(k_z))\\
M_2 =& M_0 - (G^2\cos(k_x) + \cos(k_y) + \cos(k_z)),
\end{align}
and to keep the $C_4^z \mathbb{T}$ symmetry the Hamiltonian must be of the following form:
\begin{align}
H = 
\left(\begin{matrix}
M_1 & A+iGB & 0 & C+iD \\
A-iGB & -M_1 & D-iC & 0 \\
0 & D+iC & M_2 & B+iGA \\
C-iD & 0 & B-iGA & -M_2
\end{matrix} \right) \label{Altermagnetic_Hamiltonian_3D}
\end{align}
where $C$ and $D$ must be odd with respect to the $C_4^z \mathbb{T}$ symmetry  to be preserved.
It is important to note that $C$ and $D$ are $k$-dependent functions that are off-diagonal in the Hamiltonian, coupling spin-up and spin-down states of different orbitals.
These interactions can be interpreted as an effective spin-orbit coupling, which manifests along the $z$-axis with the function $C$, and within the $xy$-plane with the function $D$.

The symmetry transformation for $C_4^z\mathbb{T}$ is given by $C_4^z\mathbb{T} (k_x,k_y,k_z) = (k_y,-k_x,-k_z)$. A suitable choice for $C$ and $D$ is as follows:
\begin{align}
C=&C_0 \sin(k_z)\\
D=&D_0 (\cos(k_x)-\cos(k_y)).
\end{align}
It turns out that in order to keep the $C_4^z \mathbb{T}$ symmetry the energy is gapless. In other words,
there is no choice of $D$ and $C$ that opens the energy gap and keeps the $C_4^z \mathbb{T}$ symmetry.

In Figure \ref{figure2}, we present the results of the 3D Hamiltonian from Eqn. \eqref{Altermagnetic_Hamiltonian_3D} under conditions of negligible SOC, with $D_0$ and $C_0$ set to zero. 
Figure \ref{figure2}a) shows one of the Weyl points (WP) in the 2D energy spectrum, specifically located within an particular $k_z$ plane. 
The positions of the all WPs in the reciprocal space are shown in Figure \ref{figure2}b), indicating that only two WPs are found along the $k_z$ axis.
In agreement with our findings, a recent study \cite{mirrorchernbandsweyl} predicts Weyl nodal lines in a 3D altermagnetic system obtained by extending a 2D altermagnetic model.

We also analyze the variation of the spin-up Chern number $c^\uparrow$, as well as the Chern number integrated over half the BZ for different $k_z$ planes.
Figures \ref{figure2}c) and d) show changes in these spin and energy topological invariants as a function of the $k_z$ planes. 
Chern numbers transitions correspond precisely to the locations of the Weyl points, suggesting that the two Weyl points exhibit opposite chiralities. 
These topological transitions, from $k_z$=$0$ to $k_z$=$\pi$, protect the WPs in the energy spectrum, thereby ensuring that the system behaves as an 3D Weyl semimetal.

To validate our results, we compute the surface states using the iterative Green’s function method in a plane parallel to the $z$-axis \cite{wanniertools}.
Figure \ref{figure2}e) shows a Fermi arc connecting the WPs in the surface energy spectrum of the $xz$ plane, providing further evidence to detect this 3D altermagnetic topological semimetal phase.
Recently, Fermi arcs have been both theoretically predicted and experimentally observed in the room-temperature 3D altermagnet candidate CrSb \cite{lu2024observationsurfacefermiarcs}.

Now, we will include the SOC into the 3D Hamiltonian described by Eqn. \eqref{Altermagnetic_Hamiltonian_3D}.
In this case, whenever the term $D_0$ is zero, the Hamiltonian preserves both $C_4^z \mathbb{T}$ and $I$.
When $D_0$ is not zero, inversion is broken and $C_4^z \mathbb{T}$ is preserved.
Here the valence bands are non-degenerate and 
therefore $I \mathbb{T}$ is not preserved. The energy bands are separated by either making $G \neq 1$ or $C+iD \neq  0 $. 

Whenever  $G=1$ the energies of the valence bands are:
\begin{align}
{\scriptstyle \lambda =   - \sqrt{\pm 2\sqrt{(A^2+ B^2)(C^2+D^2)} + A^2 + B^2 + C^2+D^2+M^2 }}
\end{align}
In this case the Hamiltonian is gapless whenever $M=0$ and $A^2+ B^2=C^2+D^2$. Unfortunately,
no matter the value of $M_0$, there are always points $(k_x,k_y,k_z)$ where the energy is zero. 
The nodal lines that appear are the intersection of the surface
\begin{align}
\cos(k_x)+\cos(k_y)+\cos(k_z) = M_0
\end{align}
with the surface
\begin{align}
A_0^2 &\sin^2(k_x) + B_0^2\sin^2(k_y)  \nonumber\\
&= C_0^2\sin^2(k_z)- D_0^2( \cos^2(k_x)-\cos^2(k_y))^2.
\end{align}

These nodal lines cannot be removed with the terms $C$ and $D$ because the expressions
that can appear must be odd with respect to the $C_4^z \mathbb{T}$ action.
The projected spin spectrum is never gapped producing a spin nodal sphere exactly when $M=0$.
In this Hamiltonian the projected spin spectrum is degenerate. To open the spin spectrum we need 
to consider $G \neq 1$,which represents the altermagnetic term.

\begin{figure}
	\includegraphics[width=8.8cm]{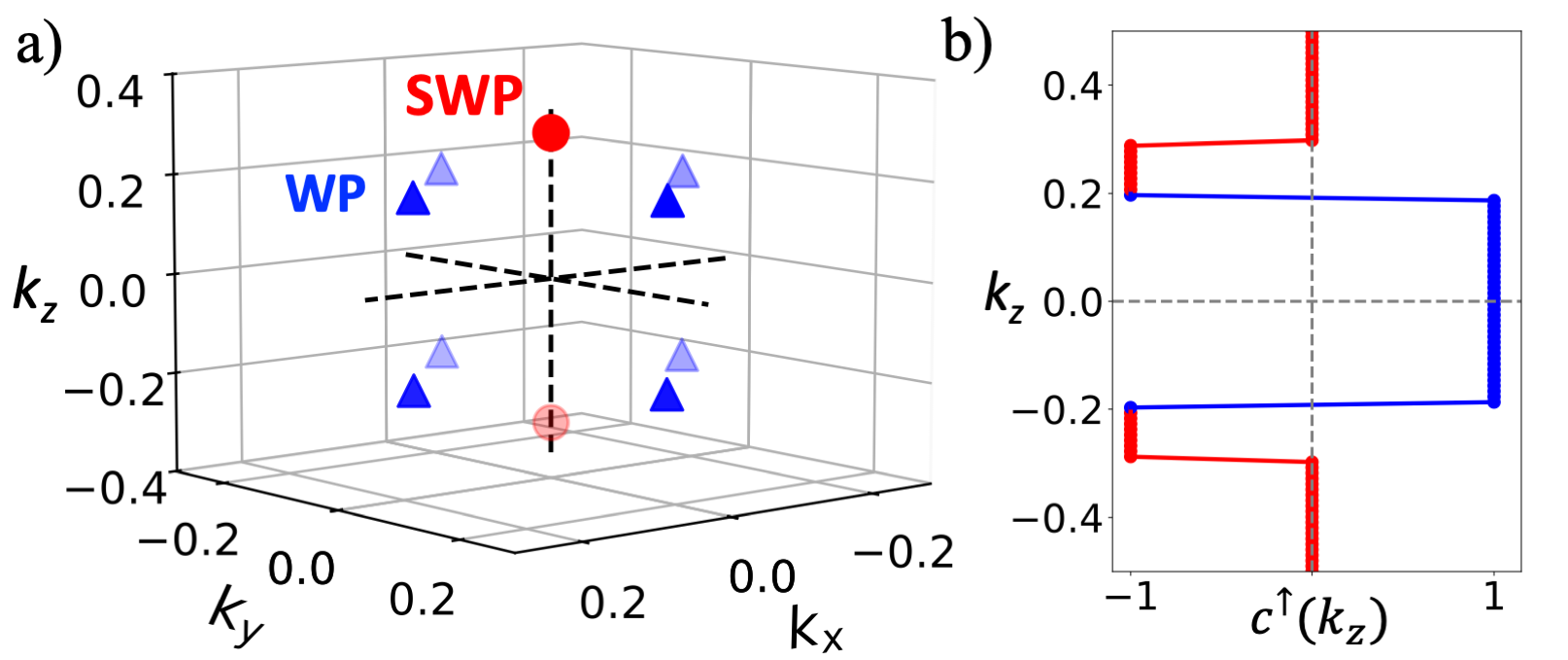}
	\caption{(a) Locations of the eight Weyl points (WP) and two spin WPs (SWP) in reciprocal space for the 3D Hamiltonian in Eqn. \eqref{Altermagnetic_Hamiltonian_3D} with $G$=1.05 and $M_0$= 1.8. SOC  parameters in the Hamiltonian are set $C_0$=1.0 and $D_0$=0.3. All the Hamiltonian parameters are in $eV$. (b) Spin-up Chern number $c^\uparrow$ as a function of the $k_z$ plane within the 3D altermagnetic phase. It is noted that the transition of the spin-up Chern number occurs when the $k_z$ planes pass through the WPs and SWPs.}
	\label{figure3}
\end{figure}

Whenever $G\neq 1$ but close to $1$ (in our case will be $G=1.05$), the energy nodal lines
transform into $8$ Weyl points, as it is shown in \ref{figure3}a), being the intersection of the surfaces 
\begin{align}
M_1=0,  \ \ M_2=0,  \   \ \ \& \ \ A^2+B^2=C^2+D^2.
\end{align}
The spin spectrum becomes non-degenerate except
on spin Weyl points\cite{Gonzalez-Uribe} (SWPs) localized along the axis $(k_x,k_y)=(0,0)$ for $G<M_0<2+G$,
$(k_x,k_y)=(0,\pi)$ for  $(k_x,k_y)=(\pi,0)$ $-2+G<M_0<2-G$ and $(k_x,k_y)=(\pi,\pi)$ for $-2-G<M_0<-G$. 
The position in $k_z$ of the spin Weyl points solve the equations $\cos(k_z)=M_0-1-G$,
$\cos(k_z)=M_0 \pm 1 \mp G$ and $\cos(k_z)=M_0+1+G$ respectively.

In Figure \ref{figure3}a) it is shown the locations of WPs and SWPs \cite{Gonzalez-Uribe} for the parameters $G = 1.05$ and $M_0 = 1.8$ in the 3D Hamiltonian described by Eqn. \eqref{Altermagnetic_Hamiltonian_3D}.  The SOC parameters in the Hamiltonian are set to $C_0$=$1.0$ and $D_0$=$0.3$, with all Hamiltonian parameters in $eV$.
Notably, a total of eight WPs and two SWPs are identified in reciprocal space, with the SWPs located along the  $k_z$ axis, as predicted.
From Figure \ref{figure3}b), it is noted that the spin Chern number changes when the $k_z$ planes cross any of the WPs and SWPs. 
In this case, the $C$ term splits the $k_z$ location of the energy and spin WPs, as is reflected by the spin-up Chern number ($c^\uparrow$) calculation as a function of the $k_z$ plane variable.

These findings suggest that the spin Chern number —and consequently the spin Hall conductivity— can be manipulated by the presence of both WPs and SWPs, representing a important indicator for the topological properties of 3D altermagnets.

\section{Topological invariants of $C_4^z \mathbb{T}$}

In this section we will determine the bulk invariants in the of the 2D and 3D torus with the action of $C_4^z \mathbb{T}$. For this
we need some results on equivariant K-theory of magnetic groups that will appear in forthcoming publications \cite{SUX, Serrano}. 

\subsection{Bulk invariants of the magnetic K-theory}

Let $G$ be the point group of symmetries of the magnetic crystal, and let $G_0 \subset G$ be the subgroup of symmetries
which behave complex linearly. The subgroup $G_0$ consists of the symmetries which do not include time reversal, while
the complement $G \backslash G_0$ are the ones that do. The quotient group $G/G_0$ is $\mathbb{Z}_2$ and the point group fits
into the short exact sequence of groups:
\begin{align}
0 \to G_0 \to G \to \mathbb{Z}_2 \to 0.
\end{align}

Let $X$ be a compact space where $G$ acts, and consider the  formal differences $E_0-E_1 $ of stable isomorphism classes of complex vector bundles $E_i \to X$
with $G$ action, such that $G_0$ acts complex linearly on the fibers of $E_i$ and $G \backslash G_0$ acts complex anti-linearly.
That is to say that $E_0-E_1$ and $E'_0-E'_1$ are equivalent if $E_0\oplus E_1'\oplus N \cong E_0'\oplus E_1 \oplus N$ for
some $N$ equivariant complex vector bundle of the same kind.
The group of equivalence classes of formal differences is denoted
\begin{align}
\mathcal{K}_{G}(X)
\end{align}
and will be called the {\it magnetic  $G$-equivariant K-theory} of $X$. The complete K-theory groups $\mathcal{K}^*_{G}(X)$
were firstly proposed by Karoubi \cite{Karoubi}, further developed and related to electronic properties of crystals by Freed and Moore \cite{FreedMoore}, and they have been extensively studied by other authors \cite{Gomi-FreedMoore, SUX, Serrano}. The definition of the higher 
magnetic equivariant K-theory groups is the standard one, namely
\begin{align}
 \mathcal{K}_G^{-q}(X) := \widetilde{\mathcal{K}}_G(\Sigma^qX)
\end{align}
where $\Sigma^qX$ is the $q$-th reduced suspension of $X$ and $\widetilde{\mathcal{K}}_G$ denotes the reduced K-theory which is the kernel of the restriction
to a marked point. Since the magnetic equivariant K-theory is $8$-periodic,
we may define the positive graded groups accordingly.

We will make an abuse of notation and we will not differentiate in the notation the case on which spin orbit coupling (SOC) is present no the one on which is absent. Whenever SOC is present, the appropriate magnetic K-theory groups need to be twisted \cite{Gomi-FreedMoore}. Alternatively, we will assume that the magnetic group
is a $\mathbb{Z}_2$ central extension of the magnetic point group.
We will simply assume that in the presence of SOC the time reversal operator $\mathbb{T}$ squares to $-1$ and all rotations
after a whole turn multiply by $-1$. The $G$ action on the fibers of the complex vector bundle $E$ need to take this information into account.

The magnetic K-theory groups $\mathcal{K}^0_{G}(X)$ are not straight forward to calculate but in some important cases they have been completely determined. In the cases of the 2D and 3D torus $T^2$ and $T^3$ respectively, in the presence of only time reversal $\mathbb{T}$ 
in the SOC environment ($\mathbb{T}^2=-1$), these groups are:
\begin{align}
\mathcal{K}^0_{\mathbb{T}}(T^2) \cong &\mathbb{Z} \oplus \mathbb{Z}_2 \\ 
\mathcal{K}^0_{\mathbb{T}}(T^3) \cong & \mathbb{Z} \oplus (\mathbb{Z}_2)^{\oplus 3} \oplus \mathbb{Z}_2.
\end{align}
The three copies of $\mathbb{Z}_2$ on the 3D torus are called the weak invariants, while the fourth copy is the 
strong invariant for topological insulators \cite{Fu-Kane-Mele}.

In order to have some understanding of the magnetic K-theory groups $\mathcal{K}^0_{G}(X)$ we may restrict 
the bundles to the action of $G_0$ and therefore we obtain a restriction map:
\begin{align}
r: \mathcal{K}^*_{G}(X) \to K_{G_0}^*(X), \ \ r(E)= E.
\end{align}
Since $G_0$ acts complex linearly, the restricted K-theory groups are the standard (twisted) equivariant complex K-theory groups\cite{Segal}.
These will be denoted with the symbol $K$ for unitary K-theory. The curly $\mathcal{K}$ will be kept for magnetic K-theory.

When applied to a point, this restriction map takes an irreducible magnetic representation of the group $G$ (or corepresentation
in Wigner's notation \cite{Wigner}) and returns the complex irreducible representations that make the magnetic one:
\begin{align}
r: \mathcal{K}^0_{G}(*) \to &K_{G_0}^0(*) & \\
V \oplus  V \mapsto & 2V \ \ & \mbox{for} & \  V \cong \widehat{V} \ \ \mbox{quaternionic type} \nonumber \\
W \oplus  \overline{W}\mapsto & W \oplus \overline{W} \ \ & \mbox{for} & \  W \ncong \widehat{W} \ \ \mbox{complex type}  \nonumber \\
U \mapsto & U \ \ & \mbox{for} & \  U \cong \widehat{U} \ \  \mbox{real type}. \nonumber
\end{align}
Here $ \widehat{U}$ denotes the conjugate representation of $U$ and it is defined as the complex conjugate representation of the conjugate $G_0$ action; see Eqn. \eqref{conjugate bundle}. The important things to note are the following \cite{Wigner}:
\begin{itemize}
\item The map that assigns an complex irreducible representation of $G_0$ to its conjugate representation (depending on who $G$ is )  
defines an involution (action of $\mathbb{Z}_2$) on  $K_{G_0}^0(*) $. 
\item The image of the restriction map lands on the $\mathbb{Z}_2$ invariant subgroup of $K_{G_0}^0(*) $.
\item Any complex irreducible representation of $G_0$ together with its conjugate can be lifted to an irreducible magnetic representation of $G$.
\end{itemize}

Therefore the induced restriction map on the $\mathbb{Z}_2$-invariant part 
\begin{align}
r: \mathcal{K}^0_{G}(*) \to K_{G_0}^0(*)^{\mathbb{Z}_2}
\end{align}
is injective and of full rank. Tensoring with $\mathbb{Q}$ we get an isomorphism of $\mathbb{Q}$-vector spaces:
\begin{align}
r \otimes \mathbb{Q}: \mathcal{K}^0_{G}(*)\otimes \mathbb{Q} \stackrel{\cong}{\to} \left( K_{G_0}^0(*)\otimes \mathbb{Q} \right)^{\mathbb{Z}_2}.
\end{align}

The previous isomorphism applied to a point can be generalized to any compact $G$-space $X$. For this we need to define the
$\mathbb{Z}_2$ action on $K_{G_0}^0(X)$.

Take any complex vector bundle $F \stackrel{p}{\to} X$ with the action of $G_0$. Let $a \in G \backslash G_0$ be any element not included in
$G_0$ and define the following complex vector bundle $\widehat{F}$. As a vector bundle
\begin{align}
\widehat{F} := a^*\overline{F} = \{(\overline{s},x)\in \overline{F}\times X | p(\overline{s})=a\cdot x \} \label{conjugate bundle}
\end{align}  
where $\overline{F}$ denotes the complex conjugate vector bundle of $F$. The $z \in \mathbb{C}$ action
on $\widehat{F}$ is the following:
\begin{align}
z \cdot (\overline{s},x) := (\overline{z \cdot s},x)
\end{align}
and the $G_0$ complex action is given by the following equation. For $g \in G_0$ the action is:
\begin{align}
g \cdot (\overline{s},x) : = \left( \overline{ (aga^{-1}) \cdot s},g \cdot x\right);
\end{align}
recall that $aga^{-1}$ is also in $G_0$
since $G_0$ is normal in $G$.
The bundle $\widehat{F}$ satisfies several properties \cite{SUX}:
\begin{itemize}
\item The isomorphism class of $\widehat{F}$ does not depend on the element $a \in G \backslash G_0$.
\item $F \cong \widehat{\widehat{F}}$, and therefore the assignment  $[F] \mapsto [\widehat{F}]$ is an involution of $K_{G_0}^0(X)$.
\item When $X$ is a point, $\widehat{V}$ denotes the conjugate representation of $V$ defined by Wigner  \cite{Wigner}.
\end{itemize}

We claim the following results\cite{SUX}:

The restriction map $r: \mathcal{K}^*_{G}(X) \to K_{G_0}^*(X)$ lands in the $\mathbb{Z}_2$-invariant subgroup:
\begin{align}
r: \mathcal{K}^*_{G}(X) \to K_{G_0}^*(X)^{\mathbb{Z}_2} \label{map r}
\end{align}
 and it induces an isomorphism when tensored with $\mathbb{Q}$:
\begin{align}
r\otimes \mathbb{Q}: \mathcal{K}^*_{G}(X)\otimes \mathbb{Q}  \stackrel{\cong}{\to} \left(K_{G_0}^*(X)\otimes \mathbb{Q} \right)^{\mathbb{Z}_2}
\label{Iso Magnetic K-theory}
\end{align}

This isomorphism permits to detect any integral invariant in the magnetic K-theory group $\mathcal{K}^*_{G}(X)$
from the complex equivariant K-theory group $K_{G_0}^*(X)$. The latter one is far simpler to calculate.

\subsection{$\mathbb{Z}_2$ bulk invariant for $C_4^z \mathbb{T}$ on 2D torus}

Denote by $\mathcal{K}^*_{C_4^z \mathbb{T}}(T^2)$
the K-theory groups of the 2D torus with the action of the group generated by $C_4^z \mathbb{T}$. Here we are assuming that
we have SOC and therefore $(C_4^z \mathbb{T})^4=-1$. We are interested in finding the bulk invariants for this magnetic
group and we will make use of the isomorphism of Eqn. \eqref{Iso Magnetic K-theory}.

Let us start understanding the magnetic K-theory groups over a point, namely the magnetic representations. 
There is only one non-trivial isomorphism class of irreducible representations of the group $\langle C_4^z \mathbb{T} \rangle$
and it is parametrized by the following $2\times 2 $ complex matrices:
\begin{align}
C_4^z \mathbb{T} \mapsto \left(
\begin{matrix}
0 & i \\
-1 & 0
\end{matrix}
\right) \mathbb{K}, \ \ \
(C_4^z \mathbb{T})^{2} \mapsto \left(
\begin{matrix}
-i& 0 \\
0 & i
\end{matrix}
\right),
 \label{C4T-irrep}
\end{align}
where $\mathbb{K}$ denotes complex conjugation; denote this representation $V$. This representation is of complex type (since the restriction to $C_2^z$ 
splits into two non-isomorphic irreducible representations), and therefore the K-theory groups of a point are:
\begin{align}
\mathcal{K}^{even}_{C_4^z \mathbb{T}}(*) \cong \mathbb{Z}, \ \ \  
\mathcal{K}^{odd}_{C_4^z \mathbb{T}}(*) \cong 0,
\end{align}
where $\mathbb{Z}$ parametrizes the number of copies of the only irreducible representation $V$ of $\langle C_4^z \mathbb{T} \rangle$.

For the group $C_2^z$ we have the following:
\begin{align}
\mathcal{K}^{even}_{C_2^z}(*) \cong \mathbb{Z} \oplus \mathbb{Z}, \ \ \  
\mathcal{K}^{odd}_{C_2^z }(*) \cong 0.
\end{align}
where $\mathbb{Z} \oplus \mathbb{Z}$ parametrizes the number of irreducible representations $L$ and $\widehat{L}$ of $C_2^z$. These
representations are 1-dimensional and the generator of $C_2^z$ acts on $L$ by multiplication by $i$ and on $\widehat{L}$ 
by $-i$. Here we need to recall that $(C_2^z)^2=-1$.

Since the only magnetic representation of $C_4^z \mathbb{T}$ is of complex type, the restriction map 
defines an isomorphism at the level of representations:
\begin{align}
\mathcal{K}^{0}_{C_4^z \mathbb{T}}(*)  &\stackrel{\cong}{\to} {K}^{0}_{C_2^z}(*)^{\mathbb{Z}_2}\\
V  &\mapsto L \oplus \widehat{L}.
\end{align}

Now, the  isomorphism of Eqn. \eqref{Iso Magnetic K-theory} 
\begin{align}
r\otimes \mathbb{Q}: \mathcal{K}^0_{C_4^z \mathbb{T}}(T^2)\otimes \mathbb{Q}  \stackrel{\cong}{\to} \left(K_{C_2^z}^0(T^2)\otimes \mathbb{Q} \right)^{\mathbb{Z}_2}
\label{Iso for C4t}
\end{align}
 implies in particular that any integer invariant of $ \mathcal{K}^0_{C_4^z \mathbb{T}}(T^2)$ could be detected with the integer
invariant in $K_{C_2^z}^0(T^2)^{\mathbb{Z}_2}$ given by the restriction map $r$. 

The bundles that generate the bulk  invariants of $K_{C_2^z}^*(T^2)$ are well known and can be constructed. Let $\mathcal{H}$ 
be the line bundle over $T^2$ with $C_2^z$ action whose Chern number is $1$. We could take as $\mathcal{H}$ the line bundle
given by the valence band of the Hamiltonian 
\begin{align}
\left(
\begin{matrix}
1.5 -\cos(k_x)-\cos(k_y) & \sin(k_x)+i \sin(k_y) \\
\sin(k_x)-i \sin(k_y)  & -(1.5 -\cos(k_x)-\cos(k_y) )
\end{matrix}
\right) \label{Hamiltonian H}
\end{align}
with $C_2^z$ action $(k_x,k_y) \mapsto (-k_x,-k_y)$ and associated matrix $\mathrm{diag}(i,-i)$. Here
$C_2^z$ has for eigenvalue $i$ on $\Gamma$ and $-i$ on the other three TRIMS. The Chern number $c_1(\mathcal{H})$ of $\mathcal{H}$ is $1$. Its conjugate $\widehat{\mathcal{H}}$ has the opposite action of $C_2^z$, its Chern number $c_1(\widehat{\mathcal{H}})$ is $-1$ and
the $C_2^z$ eigenvalues are the complex conjugate of the ones of $\mathcal{H}$.

The tensor power $\mathcal{H}^{\otimes 2n+1}$ of $\mathcal{H}$ is also a $C_2^z$-equivariant line bundle and the Chern number distinguishes  the line bundles $\mathcal{H}^{\otimes2 n+1}$ in $ K_{C_2^z}^0(T^2)$.

Nevertheless, the $\mathbb{Z}_2$-invariant elements
\begin{align}
\mathcal{H}^{\otimes 2n+1} \oplus \widehat{\mathcal{H}}^{\otimes 2n+1} \in K_{C_2^z}^0(T^2)^{\mathbb{Z}_2}
\end{align} 
become all equal in $K_{C_2^z}^0(T^2)$. Note that the Chern number
of  the sum $\mathcal{H}^{\otimes 2n+1} \oplus \widehat{\mathcal{H}}^{\otimes 2n+1}$ is zero, and they all share the same eigenvalues on the TRIMs. In other words, the bulk invariant of the bundles
$\mathcal{H}^{\otimes 2n+1} \oplus \widehat{\mathcal{H}}^{\otimes 2n+1}$
is zero, and because of this, the Chern number for each separate piece cannot be extracted from the K-theory group.

In the next section in Eqn. \eqref{K-theory C4T T2} it will be shown that
\begin{align}
\mathcal{K}^0_{C_4^z \mathbb{T}}(T^2) \cong \mathbb{Z}^{\oplus 2} \oplus \mathbb{Z}_2,
\end{align}
where the $\mathbb{Z}_2$ invariant is precisely the bulk invariant that 
is of interest.
This bulk invariant is precisely
the Chern number (mod $2$) of one of the pieces of its decomposition in the generators of $K_{C_2^z}^0(T^2)^{\mathbb{Z}_2}$. But since we cannot extract
Chern numbers from $K_{C_2^z}^0(T^2)^{\mathbb{Z}_2}$, we are going
to add the symmetry of the spin $z$ operator in order to extract 
this Chern number.

Let $S_z$ be the spin $z$  and let us assume that the spin acts on the bundles defining $ \mathcal{K}^*_{C_4^z \mathbb{T}}(T^2)$.
This happens for example when the Hamiltonian commutes with the spin. 
Incorporating $S_z$ to the group of symmetries we have that $\{S_z,C_4^z \mathbb{T} \}=0$ and that $[S_z,C_2^z]=0$. 
Now let us consider the following commutative diagram of restrictions:
\begin{align}
\xymatrix{
\mathcal{K}^0_{C_4^z \mathbb{T},S_z}(T^2) \ar[r]^r \ar[d] & K^0_{C_2^z,S_z}(T^2)^{\mathbb{Z}_2} \ar[r] \ar[d] & K^0_{S_z}(T^2)  \ar[d]^{c_1^{\uparrow}}\\
\mathcal{K}^0_{C_4^z \mathbb{T}}(T^2) \ar[r] ^r& K^0_{C_2^z}(T^2)^{\mathbb{Z}_2} & \mathbb{Z}
}  \label{diagram K-theory}
\end{align}
where the left vertical maps forget the $S_z$ action, the left horizontal maps are the restriction to the complex part 
defined in Eqn. \eqref{map r}, and the right vertical map is the Chern number of the spin-up part. Here $\mathcal{K}^0_{C_4^z \mathbb{T},S_z}(T^2)$ and $K^0_{C_2^z,S_z}(T^2)$ denote the equivariant magnetic and complex K-theories associated to the groups $\langle C_4^z \mathbb{T},S_z \rangle$ and $\langle C_2^z,S_z \rangle$ respectively.

All maps are homomorphism, and we only need to show that the generator of the integers in the upper left corner is 
obtained, and that it generates the desired elements in left lower term. For this we will make use of the altermagnetic Hamiltonian
defined before.

Denote by $E$ the vector bundle that the  valence bands of the altermagnetic Hamiltonian of Eqn. \eqref{Altermagnetic_Hamiltonian} defines. We know that for $G^2 <M_0<2$ (say $M_0=1.5$), its restriction
to the K-theory of $C_2^z$ splits as the line bundles defined by the valence bands for each Hamiltonian block; in other
words
\begin{align}
r(E) = \mathcal{H} \oplus \widehat{\mathcal{H}} \in K_{C_2^z}^0(T^2)^{\mathbb{Z}_2}.
\end{align}

Plugging $E$ in the upper left corner of diagram \eqref{diagram K-theory} we obtain:
\begin{align}
\xymatrix{
E \ar@{|->}[r]^r \ar@{|->}[d] &  \mathcal{H}_{\uparrow} \oplus \widehat{\mathcal{H}}_{\downarrow}  \ar@{|->}[r] \ar@{|->}[d] & \mathcal{H}_{\uparrow} \oplus \widehat{\mathcal{H}}_{\downarrow} \ar@{|->}[d]^{c_1^{\uparrow}}\\
E\ar@{|->}[r] ^r& \mathcal{H} \oplus \widehat{\mathcal{H}} & 1
}  \label{diagram K-theory on E}
\end{align}
where $\mathcal{H}_{\uparrow}$ denotes the line bundle generated by the eigenvector of Eqn. \eqref{spin up in H} which is all
spin-up, and $\widehat{\mathcal{H}}_{\downarrow}$ is the line bundle generated by the eigenvector of Eqn. \eqref{spin down in H}
which is all spin-down.

Now we can finish our argument. In order to detect the Chern number of one of the pieces of the class $\mathcal{H} \oplus \widehat{\mathcal{H}}$ in the 
K-theory ${K}_{C_2^z}^0(T^2)$ we use the spin $z$ splitting.

{\bf Remark:}
 {\it Note
that we cannot extract the Chern number of one of the pieces in
 $\mathcal{H} \oplus \widehat{\mathcal{H}}$ from its image in
${K}_{C_2^z}^0(T^2)$. In ${K}_{C_2^z}^0(T^2)$ the element
$\mathcal{H} \oplus \widehat{\mathcal{H}}$ has no bulk invariant.}

Incorporating the spin, we see that the vector bundles in $K^0_{C_2^z,S_z}(T^2)$
can be split into spin-up and spin-down thus having:
\begin{align}
K^0_{C_2^z,S_z}(T^2) \cong K^0_{C_2^z}(T^2)^{\uparrow} \oplus K^0_{C_2^z}(T^2)^{\downarrow}.
\end{align}
Since the $\mathbb{Z}_2$ action on vector bundles
on $K^0_{C_2^z,S_z}(T^2)$ flips the spin (this follows from the fact that
$C_4^z \mathbb{T}$ anticommutes with $S_z$), the $\mathbb{Z}_2$-invariant part can
be parametrized with the spin-up bundle:
\begin{align}
K^0_{C_2^z,S_z}(T^2)^{\mathbb{Z}_2} \cong K^0_{C_2^z}(T^2)^{\uparrow}.
\end{align}
Therefore, the incorporation of the spin into the calculation allow us
to recover a bulk integer invariant of the system.

This splitting allow us to extract only the component of the spin-up, and from
this component we may take its Chern number; this is the spin-up Chern number.
Knowing that this spin-up Chern number is odd, permits us to distinguish the
bulk invariant of the bundle in  $\mathcal{K}^0_{C_4^z \mathbb{T}}(T^2)$.

From diagrams \eqref{diagram K-theory} and  \eqref{diagram K-theory on E} we can conclude the following. The bundle $E$ generates an integer invariant
in $\mathcal{K}^0_{C_4^z \mathbb{T},S_z}(T^2)$ since it maps
to an integer invariant in ${K}^0_{C_2^z,S_z}(T^2)^{\mathbb{Z}_2}$. This
follows from the fact that the restriction map in the upper left of diagram
 \eqref{diagram K-theory} induces an isomorphims rationally.

 We now claim that the bundle $E$ generates the $\mathbb{Z}_2$-bulk
invariant in $\mathcal{K}^0_{C_4^z \mathbb{T}}(T^2)$.
For this we need to  determine explicitly  the complete K-theory invariants in
${\mathcal{K}}^0_{C_4^z \mathbb{T}}(T^2)$.  This will be done in the next section.

\subsubsection{Subdivision of the torus with 4 squares}

To find the magnetic equivariant K-theory invariants in
${\mathcal{K}}^0_{C_4^z \mathbb{T}}(T^2)$ we will make use of an equivariant
subdivision of the torus. The long exact sequence of the pairs will allows to find the desired K-theory groups.

\begin{figure}
 	\includegraphics[width=4.9cm]{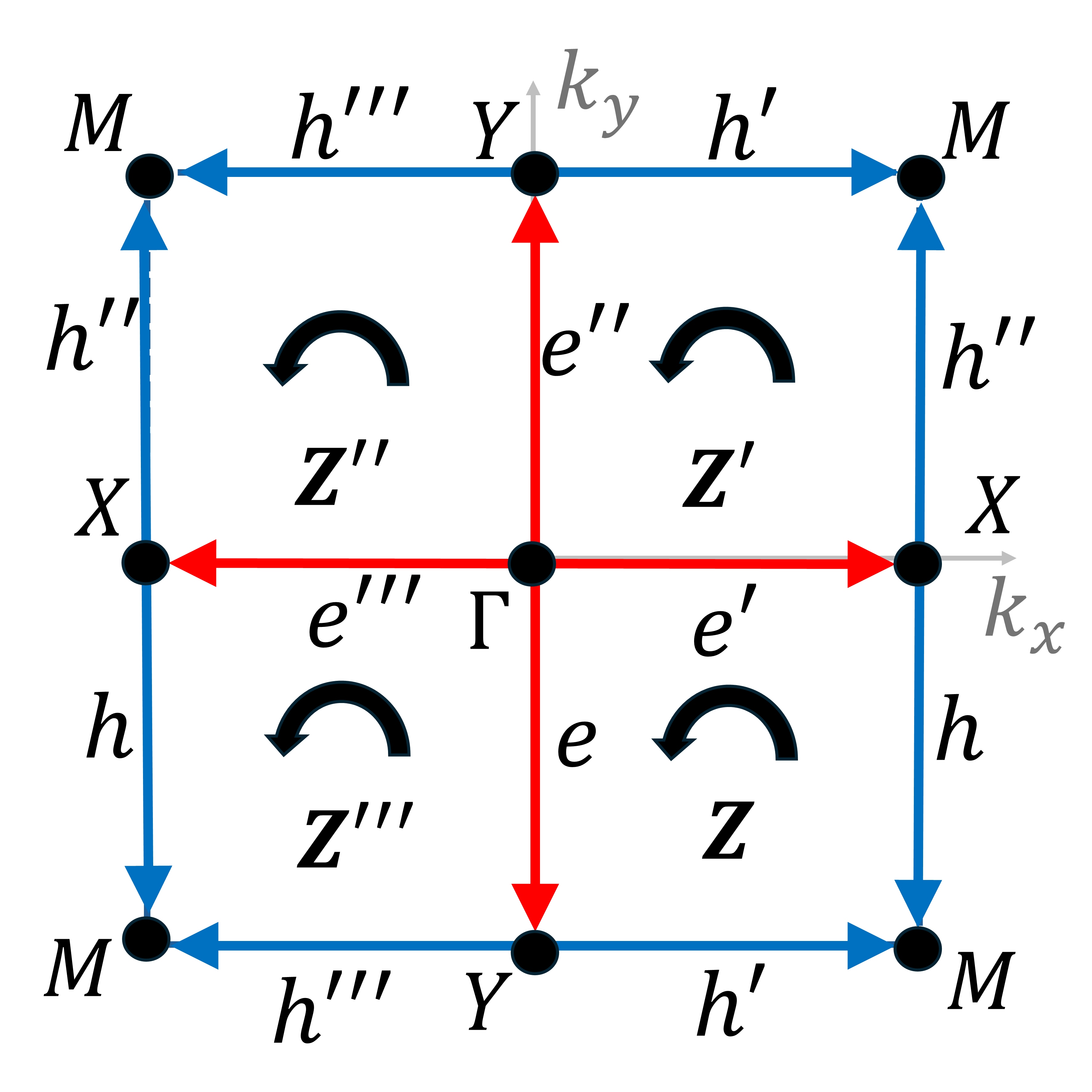}
 	\caption{Decomposition of the 2D torus with respect to its $C_4^z  \mathbb{T}$ action $(k_x,k_y) \mapsto (k_y,-k_x)$ into 4 squares.
The points $M,\Gamma$ are fixed by the four-fold rotation, while the points $X$ and $AY$ are swapped by the action. The rotation group $C_2^z$ fixes both $X$ and $Y$. The edges $e$ and $h$ are both rotated three times ($e', e'',e''',h',h'',h''')$ and the 2-cell $Z$ is copied three times $(Z',Z'',Z''')$. The U-turn arrow denotes the orientation of the 2-cells.}
 	\label{4-squares}
 \end{figure}

Let us subdivide the torus in regions compatible with the action. In Fig \ref{4-squares}
we have a partition of the torus into 0, 1 and 2 dimensional cells. Define the following 
skeletal subdivision of the torus:
\begin{align}
X_0=&\{ G,M,X,Y \} \\
X_1=&\{e,e',e'',e''',h,h',h'',h'''\} \\
X_2=&\{Z,Z',Z'',Z'''\}=T^2.
\end{align}

We have that 
\begin{align}
\mathcal{K}^{even}_{C_4^z \mathbb{T}}(X_0) \cong &  \mathbb{Z}\langle \bar{\Gamma},\bar{M},\bar{X}_+,\bar{X}_-\rangle
\end{align}
where both $ \bar{\Gamma}$ and $\bar{M}$ denote the 2-dimensional representations of $\langle C_4^z \mathbb{T} \rangle$  on $\Gamma$ and $M$ respectively and 
$\bar{X}_+$ and $\bar{X}_-$ denote the 1-dimensional representations of $C_2^z$ in $X$ (recall that $(C_2^z)^2=-1$).

From excision we know that 
\begin{align}
\mathcal{K}^{odd}_{C_4^z \mathbb{T}}(X_1,X_0) \cong &  \mathbb{Z}\langle \bar{e}, \bar{h} \rangle \\
\mathcal{K}^{even}_{C_4^z \mathbb{T}}(X_1,X_0) \cong  & 0\\
\mathcal{K}^{even}_{C_4^z \mathbb{T}}(X_2,X_1) \cong  & \mathbb{Z}\langle \bar{Z} \rangle \\
\mathcal{K}^{odd}_{C_4^z \mathbb{T}}(X_2,X_1) \cong  & 0
\end{align}
which follows from the fact that the $C_4^z \mathbb{T}$ action
on the 1 and 2-cells is free.

From the short exact sequence associated to $(X_1,X_0)$ we obtain the exact sequence:
\begin{align}
0 \to \mathcal{K}^{-2}_{C_4^z \mathbb{T}}(X_1)  \to \mathcal{K}^{-2}_{C_4^z \mathbb{T}}(X_0)   \stackrel{\delta_0}{\to} \mathcal{K}^{-1}_{C_4^z \mathbb{T}}(X_1,X_0) &\\
 \to \mathcal{K}^{-1}_{C_4^z \mathbb{T}}(X_1) & \to 0. \nonumber
\end{align}
The middle terms give the homomorphism:
\begin{align}
\mathcal{K}^{-2}_{C_4^z \mathbb{T}}(X_0)  \stackrel{\delta_0}{\to} &\mathcal{K}^{-1}_{C_4^z \mathbb{T}}(X_1,X_0)\\
\mathbb{Z}\langle \bar{\Gamma},\bar{M},\bar{X}_+,\bar{X}_-\rangle  \to & \mathbb{Z}\langle \bar{e}, \bar{h} \rangle \\
(\bar{\Gamma}, \bar{M}, \bar{X}_+,\bar{X}_-) \mapsto & (-\bar{X}_--\bar{X}_+,-\bar{X}_--\bar{X}_+)
\end{align}
from the dual of the 0-cells to the duals of the 1-cells. 
The reason $\bar{\Gamma}$ and $\bar{M}$ map to zero is the following.
We have by definition that
\begin{align}
\mathcal{K}^{-2}_{C_4^z \mathbb{T}}(X_0)  = \widetilde{\mathcal{K}}_{C_4^z \mathbb{T}}(\Sigma^2X_0) 
\end{align}
and therefore we can imagine the generators $\bar{\Gamma}, \bar{M}, \bar{X}_+,\bar{X}_-$ to be virtual bundles over $S^2$. Since the conjugation
action on $K^0(S^2)$ maps the Hopf bundle to its inverse, we have that
the restriction map $\mathcal{K}^{-2}_{C_4^z \mathbb{T}}(\{\Gamma,M \} ) \to {K}^{-2}(\{\Gamma,M \})$ is trivial. The image of $\bar{X}_+$ and $\bar{X}_-$ depends on the orientation of the 1-cells that attach to the point $X$. The 1-cell $h$ leaves from $X$ and the 1-cell $e'$ arrives, but we need
to use the automorphism that maps the information on $e'$ to $e$ by conjugation, that
on this dimension is given by multiplication by $-1$ (again the Hopf bundle is mapped to its inverse).
Therefore we have
\begin{align}
\mathcal{K}^{-2}_{C_4^z \mathbb{T}}(X_1)  \cong \mathrm{ker}(\delta_0)=\langle \bar{\Gamma},\bar{M},X_+-X_- \rangle \cong \mathbb{Z}^{\oplus 3} \\
\mathcal{K}^{-1}_{C_4^z \mathbb{T}}(X_1)  \cong \mathrm{coker}(\delta_0)=\tfrac{\mathbb{Z} \langle \bar{e}, \bar{h} \rangle}{\langle \bar{e}+\bar{h}\rangle}\cong \mathbb{Z}.
\end{align}

From the short exact sequence associated to $(X_2,X_1)$ we obtain the exact sequence:
\begin{align}
0 \to \mathcal{K}^{-1}_{C_4^z \mathbb{T}}(X_2)  \to \mathcal{K}^{-1}_{C_4^z \mathbb{T}}(X_1)   \stackrel{\delta_1}{\to} \mathcal{K}^{0}_{C_4^z \mathbb{T}}(X_2,X_1) &  \label{exact-sequence-K0}\\
  \to \mathcal{K}^{0}_{C_4^z \mathbb{T}}(X_2) \to \mathcal{K}^{0}_{C_4^z \mathbb{T}}(X_1) & {\to} 0,
\end{align}
which implies that $\delta_1: \mathbb{Z} \stackrel{\times 2}{\to} \mathbb{Z}$. 
This follows from the composition
\begin{align}
 \mathcal{K}^{-1}_{C_4^z \mathbb{T}}(X_1,X_0) \to \mathcal{K}^{-1}_{C_4^z \mathbb{T}}(X_1) &  \stackrel{\delta_1}{\to} \mathcal{K}^{0}_{C_4^z \mathbb{T}}(X_2,X_1) \\
\mathbb{Z} \langle \bar{e}, \bar{h} \rangle \to \tfrac{\mathbb{Z} \langle \bar{e}, \bar{h} \rangle}{\langle \bar{e}+\bar{h}\rangle}& \to  \mathbb{Z}\langle \bar{Z} \rangle
\end{align}
where we see that $\bar{e}$ and $\bar{h}$ get sent to $2\bar{Z}$ and $-2\bar{Z}$ respectively. The element $\bar{e}$ is mapped to $2\bar{Z}$ 
because $e$ is in the boundary of $Z$ with the correct orientation, while
$e'$ has the wrong one but the automorphism by conjugation flips the sign.
Therefore the image of $\bar{e}$ is $2 \bar{Z}$ in this dimension, and therefore
the cokernel of $\delta_1$ is our desired invariant $\mathbb{Z}_2$.

To determine $\mathcal{K}^{0}_{C_4^z \mathbb{T}}(X_1)$ we see that this group is isomorphic to the kernel of the coboundary map:
\begin{align}
\mathcal{K}^{0}_{C_4^z \mathbb{T}}(X_0)  \stackrel{}{\to} &\mathcal{K}^{1}_{C_4^z \mathbb{T}}(X_1,X_0)\\
\mathbb{Z}\langle \bar{\Gamma},\bar{M},\bar{X}_+,\bar{X}_-\rangle  \to & \mathbb{Z}\langle \bar{e}, \bar{h} \rangle \\
(\bar{\Gamma}, \bar{M}, \bar{X}_+,\bar{X}_-) \mapsto & (\bar{X}_-+\bar{X}_+-2\bar{\Gamma},2\bar{M}-\bar{X}_--\bar{X}_+),
\end{align}
which implies that 
\begin{align}
\mathcal{K}^{0}_{C_4^z \mathbb{T}}(X_1) \cong \mathbb{Z} \oplus \mathbb{Z}
\end{align}
generated by the elements $\bar{\Gamma}+\bar{M}+2 \bar{X}_+$ and $\bar{X}_--\bar{X}_+$. 

Now, since the cokernel of the coboundary map $\delta_1$ is $\mathbb{Z}_2$,
 the exact sequence of Eqn. \eqref{exact-sequence-K0} implies that the 
 K-theory groups of the torus $T^2=X_2$ are:
\begin{align}
\mathcal{K}^{0}_{C_4^z \mathbb{T}}(T^2) \cong \mathbb{Z}^{\oplus 2} \oplus \mathbb{Z}_2  \label{K-theory C4T T2}  \\
\mathcal{K}^{-1}_{C_4^z \mathbb{T}}(T^2)  \cong  0.
\end{align}

Therefore the bulk invariant of  $\mathcal{K}^{0}_{C_4^z \mathbb{T}}(T^2) $ is $\mathbb{Z}_2$, since it is the one coming from 
$\mathcal{K}^{0}_{C_4^z \mathbb{T}}(X_2,X_1)$. To see that we can detect it with the Chern number of the spin-up bundle we just need to notice that we have the commutative square
\begin{align}
\xymatrix{
\mathcal{K}^{0}_{C_4^z \mathbb{T}, S_z}(X_2,X_1) \ar[r] \ar[d]^\cong & \mathcal{K}^{0}_{C_4^z \mathbb{T}}(X_2,X_1) \ar[d]^\cong\\
\widetilde{K}^0_{S_z}(S^2) \ar[r] &{K}^0(S^2)
}
\label{restriction map for the pair}
\end{align}
where the upper horizontal map is the forgetful map, and
the bottom horizontal map is surjective because
\begin{align}
\widetilde{K}^0_{S_z}(S^2) \cong \widetilde{K}^0(S^2)^{\uparrow} \oplus
\widetilde{K}^0(S^2)^{\downarrow} 
\end{align}
maps each component isomorphically to
$\widetilde{K}^0(S^2)$. Therefore the generator of the right hand side of
Eqn. \eqref{restriction map for the pair} can be detected with the generator
of the spin-up component of the left hand side of Eqn. \eqref{restriction map for the pair}. Hence the bulk $\mathbb{Z}_2$ invariant of $\mathcal{K}^{0}_{C_4^z \mathbb{T}}(T^2) $ is detected by the parity of the Chern number of the 
spin up bundle in 
$
\mathcal{K}^{0}_{C_4^z \mathbb{T}, S_z}(X_2,X_1).
$

{\bf Theorem}: {\it The $\mathbb{Z}_2$ bulk invariant of a bundle in }
\begin{align}
{\mathcal{K}}^0_{C_4^z \mathbb{T}}(T^2) \cong \mathbb{Z}^{\oplus 2} \oplus \mathbb{Z}_2
\end{align} 
{\it can be extracted with the parity of the Chern number of the spin-up eigenbundle.}

{\bf Summarizing: } We have shown above in diagram \eqref{diagram K-theory on E} that the valence bundle $E$  of the Hamiltonian of Eqn. \eqref{Altermagnetic_Hamiltonian} has $1$ for spin-up Chern number. Therefore
this bundle $E$ realizes the $\mathbb{Z}_2$ invariant of 
${\mathcal{K}}^0_{C_4^z \mathbb{T}}(T^2)$ and therefore it is topologically
protected by this Chern number.

\subsection{Bulk invariants for $C_4^z \mathbb{T}$ on 3D torus}

A 3D $C_4^z\mathbb{T}$ altermagnet must have $\mathbb{Z}_2$ invariants (the same as the spin Chern number) on the planes $k_z=0$ and $k_z=\pi$. Whenever these numbers disagree, we say that the
altermagnet has a $\mathbb{Z}_2$ invariant. Note that this invariant being the
difference of the spin Chern numbers of the planes $k_z=\pi$ and $k_z=0$
agrees with the Chern-Simons axion insulator $\theta$ term.

Using the Mayer-Vietoris sequence associated to the open sets
\begin{align}
 U =& \{ (k_x,k_y,k_z) | -\tfrac{3\pi}{2} < k_z < \tfrac{3\pi}{2} \} \\
  V =& \{ (k_x,k_y,k_z) | \tfrac{\pi}{2} < k_z < \tfrac{5\pi}{2} \}
\end{align}
we see that equivariantly we have the homotopies
\begin{align}
U \simeq \{k_z=0\}, V \simeq \{k_z=\pi \},\\ \ U \cap V\simeq \{k_z=\tfrac{\pi}{2}\} \cup  \{k_z=-\tfrac{\pi}{2}\}.
\end{align}
Since we have an isomorphism
\begin{align}
{\mathcal{K}}^0_{C_4^z \mathbb{T}}(U \cap V) \cong {{K}}^0_{C_2^z}(\{k_z=\tfrac{\pi}{2} \}) 
\end{align}
and the restriction map 
\begin{align}
r: {\mathcal{K}}^0_{C_4^z \mathbb{T}}(T^2) \to {{K}}^0_{C_2^z}(T^2)
\end{align} 
is injective in the non-torsion part, together with the fact that
\begin{align}
{\mathcal{K}}^{-1}_{C_4^z \mathbb{T}}(U \cap V) \cong {{K}}^{-1}_{C_2^z}(\{k_z=\tfrac{\pi}{2} \}) =0,
\end{align}
 then the Mayer-Vietoris sequence
\begin{align}
0 \to {\mathcal{K}}^{0}_{C_4^z \mathbb{T}}(T^3) \to 
{\mathcal{K}}^{0}_{C_4^z \mathbb{T}}(U) \oplus 
{\mathcal{K}}^{0}_{C_4^z \mathbb{T}}(V) \to
{\mathcal{K}}^{0}_{C_4^z \mathbb{T}}(U \cap V)
\end{align}
 becomes
\begin{align}
0 \to {\mathcal{K}}^{0}_{C_4^z \mathbb{T}}(T^3) \to   
(\mathbb{Z}^{\oplus 2} \oplus \mathbb{Z}_2) \oplus 
(\mathbb{Z}^{\oplus 2} \oplus \mathbb{Z}_2) \to
\mathbb{Z}^{\oplus 2}. \label{exact sequence 3D}
\end{align}

Therefore the kernel of the right hand side homomorphism of Eqn. \eqref{exact sequence 3D} gives us
\begin{align}
 {\mathcal{K}}^{0}_{C_4^z \mathbb{T}}(T^3) \cong \mathbb{Z}^{\oplus 2} \oplus \mathbb{Z}_2 \oplus \mathbb{Z}_2
\end{align}
where the two $\mathbb{Z}_2$ components measure the $\mathbb{Z}_2$ invariant
on each of the planes $k_z=0$ and $k_z=\pi$. Whenever the two invariants differ, which means that the spin Chern number changes between the two planes, then we know that the Chern-Simons axion insulator $\theta$ term
 is not zero \cite{Gonzalez-Uribe}.

{\bf Theorem}: {\it The $\mathbb{Z}_2$ bulk invariants of a bundle in }
\begin{align}
{\mathcal{K}}^0_{C_4^z \mathbb{T}}(T^3) \cong \mathbb{Z}^{\oplus 2} \oplus \mathbb{Z}_2 \oplus \mathbb{Z}_2
\end{align} 
{\it can be extracted with the parity of the Chern number of the spin-up eigenbundle on each plane $k_z=0$ and $k_z=\pi$. The difference of these two number equals the Chern-Simons axion insulator $\theta$ term.}

\section{Conclusions}

In this study, we have identified the bulk invariants for 2D and 3D gapped Hamiltonians preserving
$C_4^z\mathbb{T}$ symmetry. For 2D materials this is a $\mathbb{Z}_2$ invariant that can be extracted as the parity of the Chern number of the spin-up bands. 
For 3D materials there are two $\mathbb{Z}_2$ invariants, one for each plane $k_z=0$ and $k_z=\pi$ respectively, and both can be extracted from the parity of the Chern number of the  spin-up bands. The difference of these two invariants is precisely the Chern-Simons axion $\theta$ term, which significantly impacts the response of these materials to external perturbations.

Additionally, we constructed an explicit $4\times 4$ Hamiltonian modeling
a 2D altermagnet which moreover realizes the non-trivial $\mathbb{Z}_2$ invariant for the $C_4^z \mathbb{T}$ symmetry.
This Hamiltonian not only establishes a prototype for an altermagnetic topological insulator but also provides a basis for exploring the electronic and spin properties of $d$-wave altermagnets.

Furthermore, we have extended this Hamiltonian to a 3D altermagnetic one preserving the $C_4^z \mathbb{T}$ symmetry. 
In the negligible SOC limit, our analysis reveals the presence of two Weyl points. 
However, when incorporating SOC, we observe the emergence of eight Weyl points and two spin Weyl points. 
In these cases, our model defines a Weyl altermagnetic semimetal, in which the spin Chern signal is linked to the locations of both the Weyl and spin Weyl points.
The presence of these Weyl points confirms that the material exhibits non-trivial topological features characteristic of a Weyl semimetal, opening pathways to explore exotic phenomena such as the chiral anomaly and surface state physics.
Finally, the formulation of a $C_4^z \mathbb{T}$ topological invariant presented in this work can be extended to other altermagnetic systems, where spin-opposite sublattices are connected through a combination of rotation and time-reversal symmetry.

\section*{Acknowledgments}
RGH gratefully acknowledges the computing time granted on the supercomputer Mogon at Johannes Gutenberg University Mainz (hpc.uni-mainz.de). 
Additionally, the support from the Universidad Nacional de Colombia (QUIPU code 202010042199) and from MinCiencias through Convocatoria 937 for Fundamental Research is also deeply appreciated.
HS acknowledges the support of CONAHCyT through grant number CVU 926934.
BU acknowledges the support of Max Planck Institute for Mathematics in Bonn, Germany. 
RGH and BU thank the continuous support of the Alexander Von Humboldt Foundation, Germany.

\bibliographystyle{naturemag}
\bibliography{topological}

\end{document}